\documentstyle[12pt,a4]{article}
\begin{document}
\hyphenation{Rijken}
\hyphenation{Nijmegen}
\hyphenation{Stich-ting Fun-da-men-teel On-der-zoek Ma-te-rie}
\hyphenation{Ne-der-land-se Or-ga-ni-sa-tie We-ten-schap-pe-lijk}

\title
{\Large\bf The Low-Energy $\mbox{\boldmath $np$}$ Scattering
   Parameters \protect\linebreak  and the Deuteron}

\author{{\bf J.J. de Swart\thanks{Electronic address: thefalg@sci.kun.nl},
    C.P.F. Terheggen}, and
   {\bf V.G.J. Stoks}\thanks{Present address: TRIUMF Theory Group,
    4004 Wesbrook Mall, Vancouver BC, Canada} \\[2mm]
  {\it Institute for Theoretical Physics, University of Nijmegen,}\\[2mm]
 {\it Nijmegen, The Netherlands}}

\maketitle

\begin{abstract}
The low-energy parameters describing the $np$ scattering in the
$^{3}S_{1}\,\!\!+\,\!\!^{3}D_{1}$ partial waves, the deuteron parameters,
and their relations are discussed. These parameters can be
determined quite accurately in the
energy-dependent Nijmegen partial-wave analyses of the $np$ data
and are also given by the high-quality Nijmegen potentials. The newest
values for these parameters are presented.
\end{abstract}

\footnotetext{Invited talk at the 3$^{\rm rd}$ International Symposium
``Dubna Deuteron 95'', Dubna, Moscow Region, Russia, July 4--7, 1995}

\section{Introduction} \label{sect.I}
The energy-dependent partial-wave analyses (PWA) of the NN scattering
data as performed in Nijmegen~\cite{93.03}
produce very good descriptions of these NN data below
$T_L=350$ MeV. We will consider explicitly the Nijm PWA93, which has $\chi^{2}/
N_{\rm d} = 0.99$ with respect to the scattering data~\cite{93.03}.
The deuteron binding
energy $B=2.224575(9)$ MeV was included in these analyses as an
experimental datum~\cite{leun82}.
Because we have an energy-dependent solution of the
scattering, we can produce accurate values for the low-energy
scattering parameters and for the deuteron parameters.
The PWA produces statistical errors on these parameters.

In Nijmegen we have also constructed high-quality (HQ)
NN potentials~\cite{93.05}.
These HQ potentials give a very good fit to the NN scattering data,
including the deuteron binding energy. For each of these
potentials we find in a direct comparison with the NN scattering
data that $\chi^{2}/N_{\rm d} = 1.03$. Because the fit with these potentials
is so good, we may speak here of alternative PWA's. Comparing the
values of the low-energy parameters and the deuteron parameters as obtained
in the various analyses gives us an estimate of the systematic error.

We can use these HQ potentials also to compute deuteron parameters
that cannot be determined from the NN scattering data,
such as the $d$-state probability $p_{d}$, the electric quadrupole moment $Q$,
and the mean-square deuteron radius $\langle r^{2} \rangle^{1/2}$.
It turns out that also these parameters seem to be almost uniquely determined.

\section{Scattering Matrix} \label{sect.II}
The wave function is a solution of the relativistic Schr\"odinger
equation
\[
(\Delta + k^{2}) \psi = 2m\, V\, \psi\ ,
\]
with $m$ the reduced mass. The relativistic relation between the cm-momentum
$k$ and the cm-energy $E$ is
\[
E = \sqrt{k^{2}+M_{p}^{2}} + \sqrt{k^{2} + M_{n}^{2}} -
  (M_{p} + M_{n})\ ,
\]
where $M_{p}$ and $M_{n}$ are the proton and neutron mass.
The asymptotic part of the wave function defines the partial-wave
$S$ matrix, which is related to the $K$ matrix by
\[
S = (1 + iK)/(1-iK)\ .
\]
For the spin-singlet waves $(\,^{1}S_{0},\,^{1}P_{1},\ldots)$ and the
uncoupled spin-triplet waves $(\,^{3}P_{0},\,^{3}P_{1},\,^{3}D_{2},\ldots)$,
there is a simple relation between the phase shift $\delta$ and the
$S$ and $K$ matrices, viz.
\[
K = \tan \delta \rule{1cm}{0mm} {\rm and} \rule{1cm}{0mm}
  S = e^{2i\delta}\ .
\]
One can show that for low momenta $\delta_{\ell} \sim k^{2\ell+1}$.

For the triplet coupled waves $(\,^{3}S_{1}\,\!\!+\!\!\,^{3}D_{1},
\,^{3}P_{2}\,\!\!+\,\!\!^{3}F_{2},\ldots)$, there are two different
parametrizations in use, the Stapp parametrization~\cite{St57}
and the Blatt and Biedenharn parametrization~\cite{BB52}.
\\[3mm]

\noindent {\bf Stapp parametrization}~\cite{St57}
\\[2mm]
The $(2\times 2)$ $S$ matrix is written in terms of the nuclear-bar phase
shifts $\bar{\delta}$ and $\bar{\varepsilon}$ as
\[
S = \left( \begin{array}{cc}
  e^{i\bar{\delta}_{J-1}} & \cdot \\
    \cdot & e^{i\bar{\delta}_{J+1}} \end{array} \right)
    \left( \begin{array}{rr}
  \cos 2\bar{\varepsilon}_{J} & i \sin 2\bar{\varepsilon}_{J} \\
  i\sin 2\bar{\varepsilon}_{J} & \cos 2\bar{\varepsilon}_{J}
            \end{array} \right)
    \left( \begin{array}{cc}
  e^{i\bar{\delta}_{J-1}} & \cdot \\
    \cdot   & e^{i\bar{\delta}_{J+1}} \end{array} \right)
\]
The advantage of this parametrization is that the behavior of the nuclear-bar
mixing angle $\bar{\varepsilon}$ for low momenta is fine,
in contradistinction with the behavior
of the Blatt and Biedenharn~\cite{BB52}
mixing angle $\varepsilon$ (see below).  For low momenta
\[
\bar{\delta}_{J-1} \sim k^{2J-1}\ , \rule{1cm}{0mm}
\bar{\delta}_{J+1} \sim k^{2J+3}\ , \rule{1cm}{0mm}
\bar{\varepsilon}_{J} \sim k^{2J+1}\ .
\]
Below the threshold $E=0$ the analyticity properties of the nuclear-bar
phases are not so nice. \\[3mm]

\noindent {\bf Eigenphases (Blatt and Biedenharn)}~\cite{BB52}
\\[2mm]
The $(2\times 2)$ $S$ matrix, and therefore also the $K$ matrix, can be
diagonalized by a real orthogonal matrix
\[
O = \left( \begin{array}{cr}
    \cos \varepsilon_{J} & -\sin \varepsilon_{J} \\
    \sin \varepsilon_{J} & \cos \varepsilon_{J} \end{array} \right)\ ,
\]
with
\[
S = O\, S_{\rm diag}\, O^{-1} \rule{1cm}{0mm} {\rm and}
   \rule{1cm}{0mm} K = O\, K_{\rm diag}\, O^{-1}\ .
\]
The eigenvalues define then the eigenphases $\delta_{J-1}$ and
$\delta_{J+1}$.
The difficulty with this parametrization is that for low momenta
\[
\delta_{J-1} \sim k^{2J-1}\ , \rule{1cm}{0mm}
\delta_{J+1} \sim k^{2J+3}\ , \ \ \ {\rm but} \ \ \
\varepsilon_{J} \sim k^{2}\ .
\]
For $J>1$ and small values of $k$ the mixing parameter $\varepsilon_{J}$
can have already
a sizeable value, when the phases are both still practically zero.
On the other hand, the analyticity properties of the eigenphases are nicer
than of the nuclear-bar phases.

\section{Analyticity} \label{sect.III}
The $S$-matrix elements are analytic functions of the complex energy $E$ or
of the complex momenta $k^{2}$. They have the so-called
{\bf unitarity cut} from $E=0$ to $\infty$ along the positive
real $E$ axis. There are additional right-hand cuts
starting at the pion-production
threshold $E_{\rm cm} \simeq 135$ MeV and left-hand cuts due to meson exchange.
The OPE cut starts at $E=-m_{\pi}^{2}/(4M) = -5$ MeV,
the TPE cut starts at $E=-m_{\pi}^{2}/M = -20$ MeV, where $M=$
the nucleon mass and $m_{\pi}=$ the pion mass. \\
When there is a bound state present the $S$-matrix elements have a pole
at the position of the bound state. In the $\,^{3}S_{1}\,\!\!+\,\!\!^{3}D_{1}$
coupled channel there is therefore the deuteron pole at $E=-2.224575(9)$
MeV.

It is useful to define the {\bf effective-range function}~\cite{Sh,SwD}
\[
M(k^{2}) = k^{\ell+1} K^{-1} k^{\ell}\ .
\]
This effective-range function is again an analytic function of $E$
or $k^{2}$ with right- and left-hand cuts as in the $S$-matrix
elements. However, the unitarity cut is removed and there is no pole
at the position of the deuteron. The effective-range function
$M(k^{2})$ is therefore regular in a circle of radius 5 MeV around the
origin, because then the OPE left-hand cut is reached. This effective-range
function can be expanded in a power series around the
origin with a radius of convergence of 5 MeV. One gets then
the {\bf effective-range expansion}
\[
M = M_{0} + M_{1}k^{2} + M_{2}k^{4} + M_{3}k^{6} + \ldots
\]
For practical purposes the radius of convergence of this effective-range
expansion is pretty small (only 5 MeV). It is possible
to define modified effective-range functions in which some of the
left-hand cuts are removed, such as the OPE cut. The radius of
convergence is then 20 MeV.

In the Nijmegen PWA we do not use the modified effective-range
expansions~\cite{BH83,Be85}
anymore, but our methods, however, are quite similar.

\section{The Deuteron in Scattering} \label{sect.IV}
The deuteron appears as a pole in the partial-wave $S$ matrix for the
coupled $^{3}S_{1}\,\!\!+\,\!\!^{3}D_{1}$ partial waves at
$k=i\alpha$. The binding energy is given by
\[
B = M_{p} + M_{n} - \sqrt{M_{p}^{2}-\alpha^{2}} - \sqrt{M_{n}^{2}-\alpha^{2}}
  = 2.224575(9)\, {\rm MeV}\ .
\]
The radius $R$ of the deuteron is then
\[
R = 1/\alpha = 4.318946\ {\rm fm}\ .
\]
Here we use relativistic kinematics. The
value of $R$ in the case of non-relativistic kinematics is significantly
different ($R_{\rm n.r.}=4.317667$ fm).

The partial-wave $S$ matrix we write in the Blatt and Biedenharn
parametrization
\[
S = O \left( \begin{array}{cc}
  S_{0} & \cdot \\
  \cdot & S_{2} \end{array} \right) O^{-1}\ .
\]
The matrix elements of the matrix $O$ are well behaved in the neighborhood
of the deuteron pole. This allows us to define at the pole
\[
\eta = - \tan \varepsilon_{1} = 0.02543(7)\ .
\]
The values quoted are from the Nijmegen 1993 PWA~\cite{93.03}.
One of the eigenvalues (we take $S_{0}$) has a pole at $k=i\alpha$.
We may write therefore
\[
S_{0} = \frac{N_{p}^{2}}{\alpha+ik} + \, \mbox{(regular function of $k$)}\ .
\]
This expression is valid in the neighborhood of the deuteron pole.
The {\bf residue} at the pole is
\[
N_{p}^{2} = 0.7830(7)\ {\rm fm}^{-1}\ .
\]
It is very convenient to define the effective range $\rho_{d} = \rho(-B,-B)$
at the deuteron pole by
\[
\rho_{d} = R - 2/N_{p}^{2} = 1.765(2)\ {\rm fm}\ .
\]
Only 3 independent deuteron parameters
$B$, $\eta$, and $\rho_{d}$ can be determined from the $np$ scattering data.

\section{The Deuteron as a Bound State} \label{sect.V}
The deuteron is a bound state in the coupled
$^{3}S_{1}\,\!\!+\,\!\!^{3}D_{1}$ two-nucleon system. Let us neglect the
possible existence of other coupled channels, such as
$\Delta\Delta$, $Q^{6}$, etc. In that case the deuteron wave function is
\[
\psi = \frac{u(r)}{r} {\cal Y}_{011}\,^{m} +
   \frac{w(r)}{r} {\cal Y}_{211}\,^{m}\ ,
\]
where ${\cal Y}_{LSJ}\,^{m}$ are the simultaneous eigenfunctions of the
operators $L^{2}$, $S^{2}$, $J^{2}$, and $J_{z}$.
We assume the wave function to be properly normalized
\[
\int_{0}^{\infty} dr \, (u^{2}+w^{2}) = 1\ .
\]
The asymptotic behavior of the wave functions for large values of $r$ are
\[
u(r) \sim A_{S} e^{-\alpha r}
  \rule{5mm}{0mm} {\rm and} \rule{5mm}{0mm}
w(r) \sim A_{D} e^{-\alpha r} \left\{ 1 + 3 \left( R/r \right) +
  3 \left( R/r \right)^{2} \right\}\ .
\]
The asymptotic normalizations are
\[
A_{S} = 0.8841\ {\rm fm}^{-1/2}
  \rule{1cm}{0mm} {\rm and} \rule{1cm}{0mm}
A_{D} = 0.0224\ {\rm fm}^{-1/2}\ ,
\]
with $N_{d}^{2} = A_{S}^{2} + A_{D}^{2} = 0.7821$ fm$^{-1}$.
The $d/s$ ratio $\eta = A_{d}/A_{s} = 0.02534$. The values quoted are from the
Nijm~I deuteron~\cite{93.05}.

For the ordinary energy-independent local potentials and for
velocity-dependent potentials one can show that $N_{d}=N_{p}$,
i.e., the normalization as defined by the wave functions is equal to the
normalization as defined by the residue~\cite{St88}.
In that case there is a trivial relation between the pole
parameters determined in scattering and the asymptotic wave function.
When we have energy-dependent potentials with
$\partial U/\partial k^{2} \neq 0$, then one can show that
\[
(N_{d}/N_{p})^{2} = 1 - \int_{0}^{\infty} dr \left( \widetilde{\psi}_{d}
  \frac{\partial U}{\partial k^{2}} \psi_{d} \right) \ .
\]

\section{Effective-Range Functions} \label{sect.VI}
The $(2\times 2)$ matrix effective-range function is $M(k^{2}) = k^{\ell+1}
K^{-1}k^{\ell}$.
This matrix effective-range function can be expanded in a power series
around the origin, but of course also around any other point in its analyticity
domain, for example around the deuteron energy.

One can do the expansion around $k^{2}=0$, then
\[
M(k^{2}) = M_{0} + M_{1}k^{2} + M_{2}k^{4} + M_{3}k^{6} + M_{4}k^{8} + \ldots\
,
\]
but one could do this also around $k^{2}=-\alpha^{2}$, then
\[
M(k^{2}) = m_{0} + m_{1}(k^{2}+\alpha^{2}) + m_{2}(k^{2}+\alpha^{2})^{2} +
   \ldots\ .
\]
This means that
\[
m_{0} = M_{0} - M_{1} \alpha^{2} + M_{2}\alpha^{4} - M_{3}\alpha^{6} +
   M_{4}\alpha^{8} - \ldots
\]
It is surprising, but also alarming, to discover that in order to get
$m_{0}$ with sufficient accuracy one needs to include the term
$M_{6}\alpha^{12}$ in the right-hand side.

When we use the Blatt and Biedenharn parametrization in
eigenphase shifts and coupling parameters, then we can define
the effective-range functions
\[
F_{J-1} (k^{2}) = k^{2J-1} \cot \delta_{J-1}\ ,\rule{4mm}{0mm}
F_{J+1} (k^{2}) = k^{2J+3} \cot \delta_{J+1}\ ,\ \ {\rm and}\ \
F_{\varepsilon} (k^{2}) = 2k^{2} \cot 2\varepsilon_{J}\ .
\]
These functions have the nice analyticity properties that make series
expansions around the origin $(k^{2}=0)$ or around the deuteron
$(k^{2}=-\alpha^{2})$ possible.

Using the Stapp parametrization we could define the effective-range functions
\[
\bar{F}_{J-1} = 2k^{2J-1} \cot 2\bar{\delta}_{J-1}\ , \rule{4mm}{0mm}
\bar{F}_{J+1} = 2k^{2J+3} \cot 2\bar{\delta}_{J+1}\ ,\ \ {\rm and}\ \
\bar{F}_{\varepsilon} = 2k^{2J+1} \cot 2\bar{\varepsilon}_{J}\ .
\]
Again, the analyticity properties of these functions are not so nice.

\section{Effective Ranges} \label{sect.VII}
It is customary to define various effective ranges such as
the standard effective range $r=\rho(0,0)$, the mixed effective range
$\rho_{m}=\rho(0,-B)$, and the deuteron effective range $\rho_{d}=\rho(-B,-B)$.
Let us look at the effective-range function
$F_{0}(k^{2}) = k \cot \delta_{0}$ for the eigenphase $\delta_{0}$, also
called the $^{3}S_{1}$ phase shift.
Expanding in a power series around the origin gives
\[
F_{0}(k^{2}) = - 1/a + {\textstyle \frac{1}{2}} r k^{2} + v_{2}k^{4} +
  v_{3}k^{6} + \ldots
\]
where $a =$ scattering length.

\begin{table}[b]
\caption{\small Low-energy scattering parameters for the
$^{3}S_{1}$ eigenphase.} \label{tab1}
\begin{center}
\renewcommand{\arraystretch}{1.2}
\begin{tabular}{l|llllll}
  & \multicolumn{1}{c}{$a$} & \multicolumn{1}{c}{$r$}
  & \multicolumn{1}{c}{$v_{2}$} & \multicolumn{1}{c}{$v_{3}$}
  & \multicolumn{1}{c}{$v_{4}$} & \multicolumn{1}{c}{$v_{5}$} \\\hline
PWA     & 5.420(1) & 1.753(2) & 0.040 & 0.672 &--3.96 & 27.1 \\[1mm]
Nijm~I  & 5.418    & 1.751    & 0.046 & 0.675 &--3.97 & 27.2 \\
Nijm~II & 5.420    & 1.753    & 0.045 & 0.673 &--3.95 & 27.0 \\
Reid93  & 5.422    & 1.755    & 0.033 & 0.671 &--3.90 & 26.7
\end{tabular}
\end{center}
\end{table}

Some of the low-energy scattering parameters are given in Table~\ref{tab1}.
Remarkable is the agreement even in the parameters $v_{2}$ to $v_{5}$.
Numerical difficulties in our calculations prevented us to calculate
accurately $v_{6}$. We think that the values for $v_{5}$ are still
accurate, but we are not sure. That there is such an enormous agreement
in these expansion coefficients comes in our opinion from our correct
treatment of the OPE. It is this OPE that determines the fast energy
dependence of the scattering matrix. The heavier-meson exchanges
give rise to a much slower energy dependence.

It is easy to see that, because $S_{0} = (1+iK_{0})/(1-iK_{0})$, one obtains
a pole at $k^{2}=-\alpha^{2}$ in the $S$-matrix elements when
$K_{0}(-\alpha^{2}) = -i$. This implies that $F_{0}(-\alpha^{2})=k K_{0}^{-1} =
(i\alpha) (i) = - 1/R$. \\
A series expansion around the deuteron is
\[
F_{0}(k^{2}) = - 1/R + {\textstyle \frac{1}{2}}
  \rho_{d} (k^{2}+\alpha^{2}) + w_{2} (k^{2} +\alpha^{2})^{2} + \ldots
\]
The mixed effective range $\rho_{m}=\rho(0,-B)$ is defined by
\[
- 1/R = - 1/a + {\textstyle \frac{1}{2}}
   \rho_{m} (-\alpha^{2})\ .
\]
Then
\[
\rho_{m} = 2R (1-R/a) = 1.754(2)(3)\ {\rm fm}\ .
\]
The first entry is the statistical error, the second one is the
systematic error. \\
We can make the expansions:
\[
\begin{array}{ccl}
r  & = & \rho_{m} + 2\alpha^{2}v_{2} - 2\alpha^{4}v_{3} + 2\alpha^{6}
  v_{4} - \ldots \\
\rho_{d} & = & \rho_{m} - 2\alpha^{2}v_{2} + 4\alpha^{4}v_{3} -
   6\alpha^{6} v_{4} - \ldots
\end{array}
\]
The terms with $v_{6}$ in the above expansions are still significant.
It turns out that
\[
r  \approx  \rho_{m} - 0.001\ {\rm fm} \rule{1cm}{0mm}{\rm and}
   \rule{1cm}{0mm} \rho_{d} \approx \rho_{m} + 0.010\ {\rm fm}\ .
\]
Using the value for $\rho_{m}$ gives
\[
r= 1.753\ {\rm fm}
  \rule{1cm}{0mm} {\rm and} \rule{1cm}{0mm}
\rho_{d}= 1.764\ {\rm fm} \ .
\]
At this point it becomes important to see how many independent low-energy
parameters and deuteron parameters there are.

The binding energy $B$ is directly related to the deuteron radius $R$.
{}From the low-energy scattering follows the scattering length $a$.
Using $R$ and $a$ gives the mixed effective range $\rho_{m}$. This
determines then the values of the effective range $r$ and
the deuteron effective range $\rho_{d}$. This in turn determines
the residue $N_{p}^{2}$ at the pole.

The $d/s$ ratio $\eta$ is also an independent quantity. We find
therefore only 3 independent quantities: $B$, $\eta$, and $a$.

\section{Multi-energy Solutions for $\mbox{\boldmath $S$}$} \label{Sect.VIII}
As already pointed out in the introduction we have 4 excellent multi-energy
parametrizations for the $S$ matrix.
The first one is our energy-dependent phase-shift analysis Nijm PWA93. We
feel that the parameters determined in this analysis~\cite{93.03}
are probably the most accurate. Then there exist 3 HQ
potentials~\cite{93.05}
(Nijm~I, Nijm~II, and Reid93) which can be considered as alternative PWA's with
a slightly higher $\chi^{2}_{\rm min}$ than Nijm PWA93. In Table~\ref{tab1}
we presented already some of the effective-range parameters.
In Table~\ref{tab2} we present the various deuteron parameters.
The deuteron parameters $\eta$ and $\rho_{d}$ can be determined
in PWA's as well as from the HQ potentials. The $d$-state probability
$p_{d}$, the mean-square radius $\langle r^{2}\rangle^{1/2}$, and the
electric quadrupole moment can only be determined for the HQ potentials.

\begin{table}
\caption{\small Deuteron parameters.} \label{tab2}
\begin{center}
\renewcommand{\arraystretch}{1.2}
\begin{tabular}{l|lllll}
& \multicolumn{1}{c}{$\eta$} & \multicolumn{1}{c}{$\rho_{d}$}
& \multicolumn{1}{c}{$p_{d}$} & \multicolumn{1}{c}{$\langle
r^{2}\rangle^{1/2}$}
 & \multicolumn{1}{c}{$Q_{0}$} \\\hline
PWA93 & 0.02543 & 1.7647 & \multicolumn{1}{c}{--} &
 \multicolumn{1}{c}{--} & \multicolumn{1}{c}{--} \\[1mm]
Nijm~I & 0.02534 & 1.7619 & 5.664 & 1.9666 & 0.2719  \\
Nijm~II & 0.02521 & 1.7642 & 5.635 & 1.9675 & 0.2707  \\
Reid93 & 0.02514 & 1.7688 & 5.699 & 1.9686 & 0.2703
\end{tabular}
\end{center}
\end{table}

\section{Conclusions} \label{Sect.IX}
In this contribution we presented the various low-energy parameters
and the deuteron parameters. We would like to stress once more that
effective-range expansions are not very accurate, unless one takes
many terms into account. A similar situation exists in the $^{1}S_{0}$ channel.
Also there the effective-range expansion is not accurate enough to describe
the data with sufficient accuracy~\cite{Be85}.
As a result, the effective-range parameters have only a very limited
usefulness. The following low-energy parameters and deuteron parameters
we would like to recommend. The errors are educated guesses for the
total error, including statistical as well as systematic errors.

\begin{itemize}
\item scattering length $a=5.4194(20)$ fm.
\item effective range $r=1.7536(25)$ fm.\\
The values of $a$ and $r$ are strongly correlated due to the
accurate value of the deuteron binding energy. A large value of $a$
implies a large value of $r$.
\item shape parameters $v_{2}=0.040(7)$ fm$^{3}$, $v_{3}=0.673(2)$ fm$^{5}$,
  $v_{4}=3.95(5)$ fm$^{7}$, and $v_{5}=27.0(3)$ fm$^{9}$.
\item deuteron binding energy $B=2.224575(9)$ MeV.
\item deuteron radius (rel) $R=4.318946$ fm.
\item $d/s$-ratio $\eta=0.0253(2)$. \\
This value is in good agreement with the recent determination by
Rodning and Knutson~\cite{Knut}
of $\eta=0.0256(4)$.
\item deuteron effective range $\rho_{d}=1.765(4)$ fm.
\item residue $N_{p}^{2}=0.7830(15)$ fm.
\item asymptotic normalizations $A_{S}=0.8845(8)$ fm$^{-1/2}$ and
$A_{D}=0.0223(2)$ fm$^{-1/2}$.
\item $d$-state probability $p_{d}=5.67(4)$\%.
\item mean-square radius $\langle r^{2} \rangle^{1/2} = 1.9676(10)$ fm. \\
This must be compared with the recent experimental
determination in atomic physics~\cite{Pa94},
which results in:
$\langle r^{2} \rangle^{1/2} = 1.971(6)$ fm. For good
discussions see Refs~\cite{Friar,Spr95}.
\item electric quadrupole moment $Q_{0}=0.271(1)$ fm$^{2}$. \\
This value disagrees with the experimental value~\cite{Bi79}
$Q=0.2859(3)$ fm$^{2}$. The difference $Q-Q_{0}\simeq 0.015$ fm$^{2}$ needs
to be explained in terms of meson exchange currents, relativistic
effects, contributions of $Q^{6}$ states, etc.
\end{itemize}

\noindent
We would like to point out here the existence of the NN-OnLine facility
of the Nijmegen group~\cite{url},
which is accessible via the World-Wide Web.
There one can obtain the various Nijmegen e-prints, the fortran codes
for some of the Nijmegen potentials, the deuteron parameters and
the deuteron wave functions, the phases obtained from the Nijmegen PWA
and from the Nijmegen potentials, and predictions for many of
the experimental quantities. A direct comparison of these predictions
with the Nijmegen NN-data base is also possible. \\[0.3cm]

\noindent
{\bf Acknowledgments} \\
We would like to thank the other members of the Nijmegen group for
their interest, help, and discussions.


\begin{thebibliography}{99}
\bibitem{93.03}
  V.G.J. Stoks, R.A.M. Klomp, M.C.M. Rentmeester, and J.J. de Swart, \\
  {\it Phys. Rev.} {\bf C 48} (1993) 792
\bibitem{leun82}
  C. van der Leun and C. Anderliesten, {\it Nucl. Phys.}
  {\bf A 380} (1982) 261
\bibitem{93.05}
  V.G.J. Stoks, R.A.M. Klomp, C.P.F. Terheggen, and J.J. de Swart, \\
  {\it Phys. Rev.} {\bf C 49} (1994) 2950
\bibitem{St57}
  H.P. Stapp, T.J. Ypsilantis, and N. Metropolis, {\it Phys. Rev.}
  {\bf 105} (1957) 302
\bibitem{BB52}
  J.M. Blatt and L.C. Biedenharn, {\it Phys. Rev.} {\bf 86} (1952) 399
\bibitem{Sh}
  M.H. Ross and G.L. Shaw, {\it Ann. Phys. (NY)} {\bf 13} (1961) 147
\bibitem{SwD}
  J.J. de Swart and C. Dullemond, {\it Ann. Phys. (NY)} {\bf 19} (1962) 458
\bibitem{BH83}
  J.R. Bergervoet, P.C. van Campen, W.M. Derks, T.A. Rijken,
  W.A. van der Sanden, J.J. de Swart, P.H. Timmers, in {\it Proceedings of
  the Conference on ``Quarks and Nuclear Structure''}, Bad Honnef, Germany,
  (ed. K. Bleuler, et al) (1983) 390
\bibitem{Be85}
  J.J. de Swart, J.R.M. Bergervoet, P.C.M. van Campen, W.A.M. van
  der Sanden, {\it Lecture Notes in Phys.}
\bibitem{St88} V.G.J. Stoks, P.C. van Campen, W. Spit, J.J. de Swart,
  {\it Phys. Rev. Lett.} {\bf 60} (1988) 1932
\bibitem{Knut}
  N.L. Rodning and L.D. Knutson, {\it Phys. Rev.} {\bf C 41}
  (1990) 898
\bibitem{Pa94}
  K. Pachucki, M. Weitz, and T.W. H\"{a}nsch, {\it Phys. Rev.}
  {\bf A 49} (1994) 2255
\bibitem{Friar}
  J.L. Friar, in {\it Few-Body Problems in Physics},
  ed. F. Gross (AIP Conference Proceedings Vol. 334, Williamsburg, 1994),
  p.\ 323
\bibitem{Spr95}
  J. Martorell, D.W.L. Sprung, D.C. Zheng, {\it Phys. Rev.}
  {\bf C 51} (1995) 1127
\bibitem{Bi79}
  D.M. Bishop, L.M. Cheung, {\it Phys. Rev.} {\bf A 20} (1979) 381
\bibitem{url}
  The URL is: {\tt http://NN-OnLine.sci.kun.nl}
\end{thebibliography}
\end{document}